# Comment on "on a new definition of quantum entropy"

In the article [1], on a new definition of quantum entropy, Campisi has explained an operator for entropy based on quantum number operator.

$$\widehat{S}(t) = \ln(\widehat{N}(t) + \widehat{1}/2) \quad (1)$$

It has been claimed that the expectation values for this operator increases for every non-quasi-static time dependent perturbations:

$$S_f \geq S_i \quad (2)$$

Where:
$$S(t) = Tr[\widehat{\rho}(t)\widehat{S}(t)] \quad (3)$$

The idea is very exiting and interesting but we have found an example for a case that entropy with this definition does not increase through a spontaneous process. Suppose that we have "n" quantum systems each of them have "l" energy levels. We assume every particle is in highest energy level and density matrix is a pure state:

$$Tr\{\rho^2\} = 1 \quad (4)$$

.and expectation value of the quantum number operator is:

$$Tr\{\widehat{N}(t_i)\widehat{\rho}(t_i)\} = l \quad (5)$$

As time passes particles spontaneously relax to the lower levels. Density matrix evolves to a mix state[2]:

$$\widehat{\rho}(t) = \sum_{k=0}^{k=l} |b_k(t)|^2 |k,t\rangle\langle k,t| \quad (6)$$

$$\sum_{k=0}^{k=l} |b_k(t)|^2 = 1 \quad (7)$$

Therefore the expectation value for N will be given with the following equilibrium:

$$\langle \widehat{N}(t_f)\rangle = \sum_{k=0}^{k=l} k|b_k(t_f)|^2 \quad (8)$$

According to inequalities if we have a set of n non-negative set of numbers a say

$$a_1, a_2, \ldots, a_n \quad (9)$$

For M (a) weighted mean value of the set we have[3]:

Min (a) $\leq$ M (a) $\leq$ Max (a) $\quad (10)$

From equations 7, 8, 10 we have:

$$\langle \widehat{N}(t_f)\rangle \leq l \leq \langle \widehat{N}(t_i)\rangle \quad (11)$$

We have shown that
$$\exp(S_i) \geq \exp(S_f) \quad (12)$$

Because logarithm is a monotonic function it is clear that:
$$S_i \geq S_f \quad (13)$$

In this example we see if entropy is defined with (1) in spontaneous emission entropy decreases, so we think this definition for entropy is not convenient.

K. Sadri
k1.sadri@gmail.com
Department of Chemistry,
Sharif University of Technology,
Tehran, Iran

[1] M. Campisi, on a new definition of quantum entropy, Arxiv:o803.0282v1 [quant-ph]

[2] I. Levine, Molecular spectroscopy (John Wiley & sons, 1975)

[3] G.H. Hardy and J. E. Littlewood and G. Polya, Inequalities second edition (Cambridge University Press, 1989)